\documentclass[twocolumn,superscriptaddress,floatfix,preprintnumbers,nofootinbib,prd,10pt]{revtex4-2}
\pdfoutput=1
\usepackage[colorlinks=true,breaklinks=true]{hyperref}
\usepackage[normalem]{ulem}
\usepackage[utf8]{inputenc}
\hypersetup{allcolors=[rgb]{0.0 0.0 0.6},linkcolor=[rgb]{0.75 0.05 0.05}}
\usepackage{amsmath,amssymb}
\usepackage{epsfig}  
\usepackage{graphicx}
\usepackage{appendix}

\usepackage{epsfig}  
\usepackage{graphicx}   
\usepackage{slashed}       
\usepackage{tikz}
\usepackage{url}
\usepackage{color}
\usepackage{multirow}
\usepackage{comment}
\usepackage{amssymb}
\usepackage[capitalize]{cleveref}
\usepackage{bm}
\usepackage{soul}

\DeclareMathOperator{\cm}{cm}
\DeclareMathOperator{\GeV}{GeV}

\DeclareMathOperator{\MeV}{MeV}

\DeclareMathOperator{\TeV}{TeV}
\DeclareMathOperator{\s}{s}

\DeclareMathOperator{\Mpc}{Mpc}

\newcommand{\beq}{\begin{equation}}
\newcommand{\eeq}{\end{equation}}

\newcommand{\gag}{g_{a\gamma}}
\newcommand{\dSN}{R_{\text{SN}}}
\newcommand{\diff}{\mathrm{d}}
\newcommand{\alpSpectrum}{\frac{\diff N_a}{\diff \omega_a}}

\newcommand{\lat}[0]{\textit{Fermi}-LAT}

\allowdisplaybreaks

\bibpunct{[}{]}{,}{n}{}{,}

\begin{document}

\preprint{LAPTH-038/23}

\title{Constraining MeV-scale axion-like particles \\with {\it Fermi}-LAT observations of SN 2023ixf}

\author{Eike Ravensburg}
\email{eike.ravensburg@fysik.su.se}
\affiliation{The Oskar Klein Centre, Department of Physics, Stockholm University, Stockholm 106 91, Sweden
}
\author{Pierluca~Carenza}\email{pierluca.carenza@fysik.su.se}
\affiliation{The Oskar Klein Centre, Department of Physics, Stockholm University, Stockholm 106 91, Sweden
}
\author{Christopher Eckner}\email{eckner@lapth.cnrs.fr}
\affiliation{LAPTh \& LAPP, CNRS,  USMB, F-74940 Annecy, France}
\author{Ariel~Goobar}\email{ariel@fysik.su.se}
\affiliation{The Oskar Klein Centre, Department of Physics, Stockholm University, Stockholm 106 91, Sweden
}

\date{\today}
\smallskip

\begin{abstract}
The {\it Fermi}-LAT observations of SN 2023ixf, a Type II supernova in the nearby Pinwheel Galaxy, Messier 101 (M101), presents us with  an excellent opportunity to constrain MeV-scale Axion-Like Particles (ALPs). By examining the photon decay signature from heavy ALPs that could be produced in the explosion, the existing constraints on the ALP-photon coupling can be improved, under optimistic assumptions, by up to a factor of $ \sim 2 $ for masses $ m_a \lesssim 3 \MeV  $. Under very conservative assumptions, we find a bound that is slightly weaker than the existing ones for $ m_a \lesssim 0.5$~MeV. The exact reach of these searches depends mostly on properties of the SN progenitor.
This study demonstrates the relevance of core-collapse supernovae, also beyond the Magellanic Clouds, as probes of fundamental physics.
\end{abstract}

\maketitle

\section{Introduction}
The explosion of core-collapse supernova (SN) 2023ixf in the nearby galaxy M101 has been followed over the entire electromagnetic spectrum by the astronomical community: at radio wavelengths~\cite{2023ATel16052....1C,2023TNSAN.146....1M,Berger:2023jcl}, infrared~\cite{Jencson:2023bxz,Soraisam:2023ktz,Teja:2023hcm,Yamanaka:2023gbr}, optical~\cite{Jacobson-Galan:2023ohh,Yamanaka:2023gbr,Bostroem:2023dvn,2023ATel16046....1V,2023TNSAN.129....1K,2023ATel16053....1S,2023ATel16054....1S,2023ATel16057....1B,2023ATel16067....1S,2023ATel16042....1S} and X-rays~\cite{Grefenstette:2023dka,Xray,Xray2,2023ATel16065....1M} (see also Ref.~\cite{Neustadt:2023fao}). No prompt neutrinos were observed by neither Super-Kamiokande nor IceCube~\cite{2023ATel16043....1T,2023ATel16070....1N}, as expected given the distance to M101, $ \dSN = 6.85\pm0.15$~Mpc~\cite{Riess:2021jrx}. The earliest optical observations of the explosion are from May 18th, between 19:30 and 20:30 UT~\cite{2023TNSAN.130....1M}, leaving an uncertainty of about one hour on the onset of the electromagnetic signal following the core-collapse.
After some unsuccessful searches for the progenitor of this SN~\cite{2023ATel16060....1M,2023ATel16064....1B}, recent studies using Hubble Space Telescope images found a candidate progenitor of $\sim(11\pm2)~M_{\odot}$~\cite{2023TNSAN.139....1S,Pledger:2023ick,Kilpatrick:2023pse} with a radius of $ R_* = (410 \pm 10) R_\odot $~\cite{Hosseinzadeh:2023ixa}, where $M_{\odot}$ and $R_{\odot}$ are respectively the solar mass and radius. All this information is composing a coherent description of this event, roughly in agreement with theoretical expectations.

The close proximity and high luminosity of SN 2023ixf makes it an exceptional case study for investigating the later phases of massive stars, the physics behind SN explosions, and the enrichment of the surrounding circumstellar environment. Additionally, core-collapse SN explosions hold significant importance in the realm of particle physics due to their potential to probe exotic particles~\cite{Raffelt:1996wa}.
Of particular interest are MeV-scale Axion-Like Particles (ALPs), which have gained considerable attention for their potential influence on various phenomena.
Besides their widely discussed impact on SNe~\cite{Carenza:2020zil,Caputo:2021kcv,Caputo:2022mah}, such ALPs have also been extensively studied for their effects on Big Bang Nucleosynthesis and the Cosmic Microwave Background~\cite{Cadamuro:2010cz,Cadamuro:2011fd,Depta:2020wmr,Balazs:2022tjl}, and on the evolution of low-mass stars~\cite{Raffelt:1987yu,Carenza:2020zil,Dolan:2021rya,Lucente:2022wai}, among other areas. Furthermore, the MeV mass range can be explored through collider and beam-dump experiments~\cite{Jaeckel:2015jla,Dolan:2017osp,Dobrich:2019dxc,Banerjee:2020fue}.

Being among the closest core-collapse supernovae in many years, SN 2023ixf is a promising candidate as a natural factory of ALPs. However, at a distance $\sim100$ times larger than SN 1987A, the lack of detection of the associated neutrino burst prevents us from reconstructing the energy emitted by the SN and applying a cooling argument~\cite{Raffelt:1990yz}. In addition, since the exact time of the collapse of the stellar core is not known, this source cannot be used as probe of light ALPs. Searches for light ALPs emitted by a SN and converting into gamma rays in the Galactic magnetic field~\cite{Payez:2014xsa,Meyer:2016wrm,Meyer:2020vzy,Crnogorcevic:2021wyj,Hoof:2022xbe,Calore:2023srn} require a precise timing to observe the hypothetical ALP-induced signal in coincidence with the SN neutrino burst. 

\begin{figure*}[t]
\includegraphics[width=0.85\linewidth]{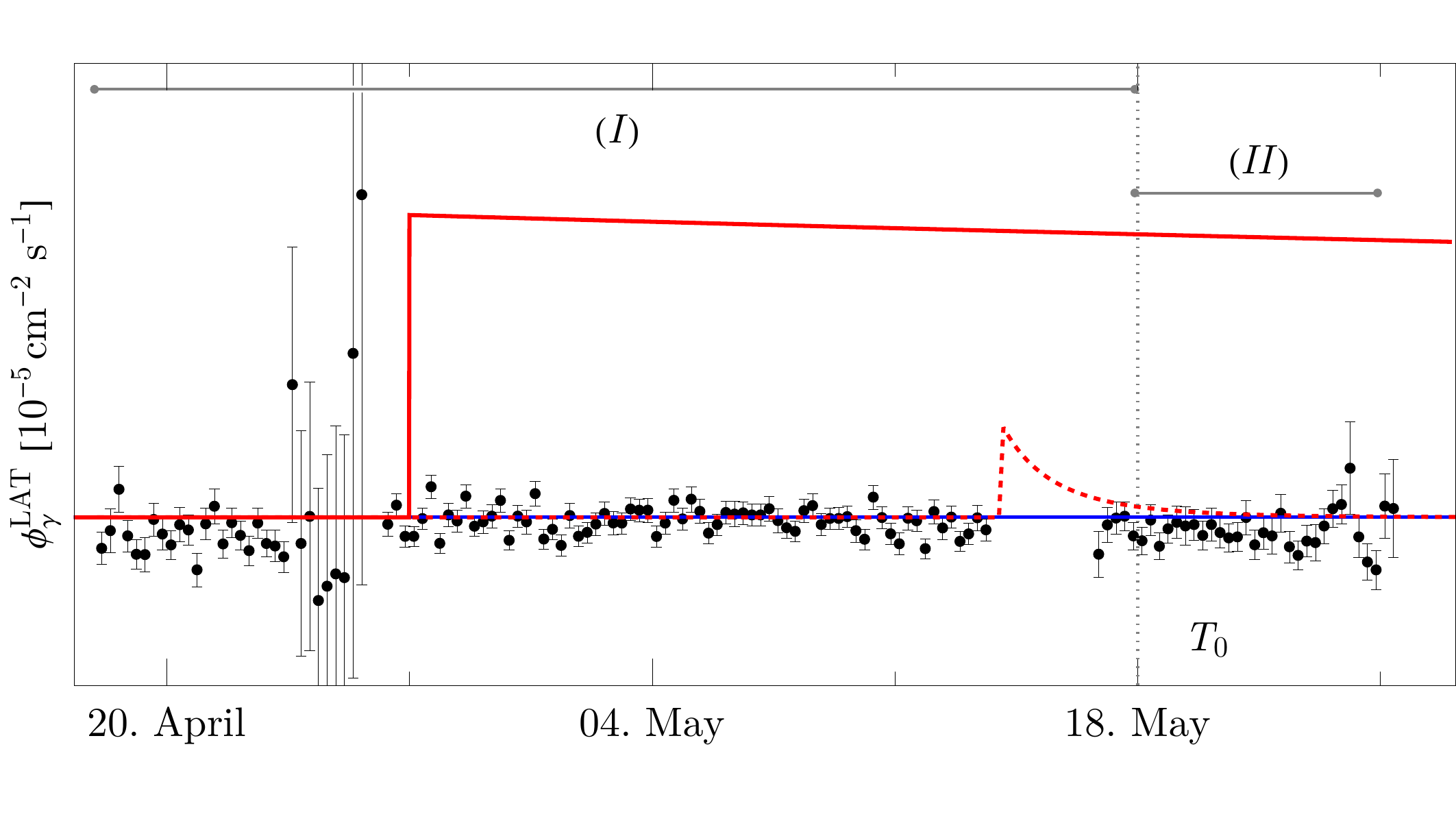}
    \caption{Photon flux measured by \textit{Fermi}-LAT in the direction of SN 2023ixf following the selection criteria detailed in the main text and binned into intervals of 6 hours (black dots and error bars) for the time interval from one month before the onset of the optical signal from SN 2023ixf up to one week after. The error bars reflect the statistical uncertainty of the photon flux; the latter we obtained by dividing the number of counts by the LAT exposure for each 6h interval, respectively. The horizontal blue line indicates the average background level $\hat{b}$ derived from the full \textit{Fermi}-LAT data set of more than 14 years of observations.
    Time intervals \emph{(I)} and \emph{(II)} as described in the text are marked by horizontal gray lines.
    As red lines we show the background plus two potential signals induced by ALPs with $ \gag = 2.2 \times 10^{-10} \GeV^{-1}, \, m_a = 0.03 \MeV $ (solid), or $ \gag = 1.0 \times 10^{-11} \GeV^{-1} , \, m_a = 8.09 \MeV $ (dashed), respectively. Note that the on-set time of any potential ALP signal is approximately equal to the unknown core-collapse time; here, we chose two different on-set times for illustration purposes.
    The dotted vertical line shows $ T_0 $, the onset of the optical light from SN 2023ixf, and date-labels denote 19:30:00 UTC of the respective day.
    }
    \label{fig:LAT_lightcurve}
\end{figure*}

Instead, here we use SN 2023ixf to probe heavy ALPs, thermally produced in the SN core and decaying into photons~\cite{Jaeckel:2017tud,Hoof:2022xbe,Muller:2023vjm,Diamond:2023cto} (see also Ref.~\cite{Dev:2023hax} for another recent study of ALP constraints from extragalactic sources, in this case the neutron star merger GW170817). For that case, the ALP-induced signal is expected to arrive at Earth with a significant time delay and persist, nearly constant in time, for days or even weeks. This scenario can be explored by the {\it Fermi} Large Area Telescope (LAT), that observed in direction of M101 for several days before and after the detection of the optical signal~\cite{Fermi}.
In \cref{fig:LAT_lightcurve} we report the gamma-ray observations in terms of reconstructed photon flux as function of the observation time. The beginning of the optical signal is marked by a vertical dotted line. In particular, the observations allow us to set constraints on MeV-scale ALPs, improving the existing SN 1987A bound on the ALP-photon coupling~\cite{Muller:2023vjm} by up to a factor $ \sim 2 $ for masses $ m_a \lesssim 3 \MeV $. We estimate that this constraint can be weakened up to a factor of 3 with pessimistic assumptions about the uncertain stellar properties. In this conservative scenario, our findings corroborate the ALP constraints from low mass stars~\cite{Ayala:2014pea,Carenza:2020zil,Lucente:2022wai} and the diffuse gamma-ray flux from SN-ALP decays~\cite{Caputo:2022mah}.

In Sec.~\ref{sec:production} we briefly discuss the ALP production mechanisms in SN 2023ixf and we anticipate that the most important one in this case is Primakoff conversion of thermal photons into ALPs in the electrostatic field of ions. In Sec.~\ref{sec:fluence} we summarize how to compute the ALP-induced gamma-ray flux in coincidence with SN 2023ixf, pointing the interested reader to Ref.~\cite{Muller:2023vjm} for more details. The photon background, against which the ALP-related signal has to be compared, is discussed in Sec.~\ref{sec:dataAnalysis}. This discussion paves the way to set the SN 2023ixf constraint on massive ALPs, as explained in Sec.~\ref{sec:constraint}. Our results and their consequences are summarized in Sec.~\ref{sec:conclusions}.

\section{ALP production in supernovae}
\label{sec:production}
The ALP-photon interaction is characterized by the Lagrangian term~\cite{Raffelt:1987im}
\begin{equation}
		{\cal L}_{a\gamma}=-\frac{1}{4} \, g_{a\gamma}
		F_{\mu\nu}\tilde{F}^{\mu\nu}a\,,
    \label{eq:lagrangian}
\end{equation}
with $g_{a\gamma}$ the ALP-photon coupling, $F_{\mu\nu}$ the electromagnetic field strength tensor, $\tilde{F}^{\mu\nu}$ its dual and $a$ the ALP field. This interaction causes the production of ALPs in the hot and dense plasma in a SN core through Primakoff conversion and photon coalescence~\cite{Raffelt:1985nk,DiLella:2000dn}.
The Primakoff process, the conversion of a photon into an ALP in the electrostatic field of ions in the plasma, leads to the following matrix element, summed over polarizations, for the conversion process
\begin{equation}
\begin{split}
    |\mathcal{M}|^{2}
    = \gag^2 (Z e)^2 \, 8 m_{Ze}^2 \frac{1 - \cos^2 \phi}{\frac{k}{p_a} + \frac{p_a}{k} + 2 \cos\phi}\,,
\end{split}
\end{equation}
where $Ze$ is the charge of the target ion, $m_{Ze}$ is its mass, $ p_a $ is the momentum of the ALP and $\omega_a$ its energy, $k$ is the momentum of the converted photon, $\phi$ is the angle between these momenta, and energy conservation requires $\omega_{a} = \sqrt{p_a^2 + m_a^2} = \sqrt{k^2 + \omega_{\rm pl}^2}$, where $\omega_{\rm pl}$ is the plasma frequency, since we neglect the recoil of the target ion (or equivalently, send $m_{Ze} \to \infty$). 
The resulting spectral rate of change of the ALP number density is~\cite{DiLella:2000dn,Carenza:2020zil,Lucente:2020whw}
\begin{equation} \label{eq:PrimakoffSpectralProductionRate}
\begin{split}
    & \dfrac{\diff^2 n_a}{\diff t_{\text{pb}}\, \diff\omega_{a}}\\
    &=\Bigg[\prod_{i} \int \frac{\mathrm{d}^3 \vec{p}_i}{(2\pi)^3 2 E_i} f_i(E_i) \Bigg]
    \int \frac{\mathrm{d}^3 \vec{p}'_{Ze}}{(2\pi)^3 2 E'_{Ze}} \left[ 1 \pm f_{Ze}(E'_{Ze}) \right]\\
    &\qquad \times (2\pi)^4 \delta^{(4)} \Bigg( \sum_i P_i - P'_{Ze} - P'_a \Bigg)
    \frac{\lvert\vec{p}'_a\rvert}{4\pi^2} \lvert \mathcal{M} \rvert^2 \\
    &=\gag^2 \dfrac{T\kappa_s^2}{32\pi^{3}} \frac{k \, p_a}{e^{\omega_a / T} - 1}\\
    &\Biggl\{\dfrac{\left[\left(k+p_a\right)^2+\kappa_s^2\right]\left[\left(k-p_a\right)^2+\kappa_s^2\right]}{4kp_a\kappa_s^2}\ln\left[\dfrac{(k+p_a)^2+\kappa_s^2}{(k-p_a)^2+\kappa_s^2}\right]\\
    &\quad\quad - \dfrac{\left(k^2-p_a^2\right)^2}{4kp_a\kappa_s^2}\ln\left[\dfrac{(k+p_a)^2}{(k-p_a)^2}\right]-1 \Biggr\}\,,
\end{split}
\end{equation}
where, in the first line, $ \vec{p}_i $, $ E_i $, $ P_i $ are the three-momentum, energy, and four-momentum of the particles in the initial state (a photon and an ion) and the corresponding quantities for the final state particles (an ALP and an ion) are denoted by a prime, $f_{i,Ze} $ are the distribution functions of the respective particles, $ t_{\rm pb} $ is the time after the SN core bounce measured in the local SN frame, $T$ is the temperature of the plasma, and $\kappa_s^2 = e^2 n_p^{\rm eff} / T$ is the screening scale for a degenerate nucleon gas~\cite{Lucente:2020whw,Payez:2014xsa}. This process mainly takes place in the electric field of protons since electrons are strongly degenerate~\cite{Payez:2014xsa}.
More massive ALPs, with $ m_a \geq 2 \omega_{\text{pl}} \sim \mathcal{O}(25 \text{ MeV}) $, can also be efficiently produced by the inverse decay or photon coalescence $ \gamma \gamma \to a $~\cite{Raffelt:1985nk,DiLella:2000dn,Carenza:2020zil,Lucente:2020whw}. This process becomes competitive with the Primakoff production only for very massive ALPs, $m_{a}\gtrsim70$~MeV. In the following we will be interested only in lighter ALPs and can hence neglect this process.

The ALP spectrum $ \alpSpectrum $ is the volume and time-integral of the Primakoff production rate in \cref{eq:PrimakoffSpectralProductionRate} over a SN model \begin{equation}
\begin{split}
    \alpSpectrum(\omega_a) &= 4\pi \int \diff t_{\text{pb}} \int \diff r \, r^2 \, \ell^{-1}(r,t_{\text{pb}}) \\
    &\qquad \times \dfrac{\diff^2 n_a}{\diff t_{\text{pb}}\,\diff\omega_{a}^{\text{loc}}}(r,t_{\text{pb}}, \omega_a^{\text{loc}}) \, ,
\end{split}
\label{eq:integrated}
\end{equation}
where the ALP energy for an observer far away from the SN is red-shifted compared to the local energy with which it is produced in the SN core $ \omega_a^{\text{loc}} = \ell^{-1}(r,t_{\text{pb}}) \, \omega_a $; here, $ \ell(r,t_{\text{pb}}) $ is the lapse function which incorporates gravitational corrections on the ALP spectrum.
The integration of \cref{eq:integrated} was performed in Ref.~\cite{Calore:2021hhn} for different SN progenitor masses $M$, obtaining an analytical fit for the ALP spectrum and its uncertainties
\begin{equation}
	\frac{\diff N_{\rm a}}{\diff \omega_{a}} = C \left(\frac{g_{a\gamma}}{10^{-12}\,\GeV^{-1}}\right)^{2}
	\left(\frac{\omega_{a}}{\omega_{0}}\right)^\beta \exp\left( -\frac{(\beta + 1) \omega_{a}}{\omega_{0}}\right) \,,
	\label{eq:time-int-spec}
\end{equation}
where 
\begin{equation}
  \begin{split}
      &\frac{C(M)}{10^{48}\,\MeV^{-1}} =(1.73\pm0.172)\frac{M}{M_{\odot}}-9.74\pm2.92\,,\\
      &\frac{\omega_0 (M)}{{\rm MeV}} =(1.77\pm0.156)\frac{M}{M_\odot}+59.3\pm2.65\,,\\
      &\beta(M) = (-0.0254\pm 0.00587)\frac{M}{M_\odot} + 2.94 \pm 0.0997\,\,\,.
  \end{split}  
  \label{eq:parameters}
\end{equation}
The expression above describes a \emph{quasi-thermal} spectrum, with index $\beta$ (in particular, $\beta=2$ corresponds to a perfectly thermal spectrum of ultrarelativistic particles) and $ \omega_0 $ the average energy of an ultrarelativistic ALP. The assumed linear ansatz for the parameters in Eq.~\eqref{eq:parameters} as function of the SN progenitor mass does not have any particular physical motivation. Due to the scarcity of points used to derive these relations, the linear behavior was chosen in Ref.~\cite{Calore:2021hhn} only for simplicity.
Therefore, by using Eqs.~\eqref{eq:time-int-spec}-\eqref{eq:parameters}  we can derive the predicted ALP emission spectrum and associated uncertainties for SN 2023ixf, assuming a $11\pm2\,M_{\odot}$ progenitor star. 

\section{The gamma-ray fluence from decaying ALPs} \label{sec:fluence}
A large amount of ALPs produced in the SN core escape from the star because of their weak interactions with ordinary matter. When traveling outside of the stellar photosphere, ALPs decay producing gamma rays, possibly reaching our detectors. The number of such photons that would be observable per unit area is the fluence $F_{\gamma}$.
Any given ALP with energy $\omega_a$ decays at a random distance $L$ to the SN, with an exponentially distributed probability with decay length $\ell_a$, and at a random angle $\alpha$ between its trajectory and that of the photon reaching earth, the probability distribution of which can be determined by a Lorentz boost applied to the isotropic rest-frame decay angle distribution \cite{Landau:1975pou}. Thus, we can write the fluence as a differential with respect to these three quantities that uniquely determine the ALP-photon trajectory \cite{Muller:2023vjm}
\begin{equation}
\begin{aligned}
    \frac{\diff^3 F_\gamma}{\diff \omega_a \, \diff c_\alpha \, \diff L}
    &= \frac{1}{4\pi \, \dSN^2} \alpSpectrum \frac{\omega_a^2 - p_a^2}{(\omega_a - c_\alpha p_a)^2} \frac{e^{-L/\ell_a(\omega_a)}}{\ell_a(\omega_a)}\\
    &\quad \times \Theta_{\text{cons.}}(\omega_a, c_\alpha, L) \, ,
\end{aligned}
\end{equation}
where $c_\alpha \equiv \cos(\alpha)$, the ALP emission spectrum $\alpSpectrum$ is discussed in \cref{sec:production}, and $\Theta_{\rm cons.}$ implements geometrical and observational constraints that have to be fulfilled for the photon to be observable, see Ref.~\cite{Muller:2023vjm} for details. As we showed there, the ALP-centered variables $(\omega_a, c_\alpha, L)$ can be uniquely mapped onto directly observable quantities:
the energy $\omega_{\gamma}$ of the produced photon, its time delay $t$ compared to the first neutrino\footnote{We will extensively discuss the method of analysis in the case that there is no neutrino observation and this time delay cannot be determined.}, and the cosine of the observation angle $c_{\theta}$. Using this mapping, we get an alternative expression for the differential fluence~\cite{Muller:2023vjm}
\begin{equation} \label{eq:fluence}
\begin{aligned}
    &\frac{\diff^3 F_\gamma}{\diff \omega_\gamma \, \diff t \, \diff c_\theta}
    = \frac{2}{\tau_a} \frac{|c_\theta|}{(t/\dSN + 1 - c_\theta)^2} \frac{\alpSpectrum(\omega_a(\omega_\gamma, t, c_\theta))}{4\pi \, \dSN^2} \\
    &\quad \times \frac{m_a}{p_a(\omega_\gamma, t, c_\theta)} \exp\left[ -\frac{\dSN}{\tau_a} \frac{2 \omega_\gamma}{m_a} \left(\frac{t}{\dSN} + 1 - c_\theta\right) \right] \\
    &\quad \times \Theta_{\text{cons.}}(\omega_\gamma, t, c_\theta) \, ,
\end{aligned}
\end{equation}
where the lifetime of the ALP is $ \tau_a = \frac{64\pi}{\gag^2 m_a^3} $.
Geometric constraints and the requirement that ALPs decay outside the SN photosphere, i.e., at a distance larger than $R_{*} $ (see \cref{tab:paramValues}) from the production point, restrict the integration domain for the variables $(\omega_{\gamma},t,c_{\theta})$ as dictated by~\cite{Muller:2023vjm}
\begin{equation}
\begin{aligned}
    &\Theta_{\text{cons.}}(\omega_\gamma, t, c_\theta) \\
    &= \Theta\left(c_\theta - \left[1 - \frac{t}{\dSN}\left( 
\frac{m_a}{2\omega_\gamma} \sqrt{\frac{2\dSN}{t}+1} - 1 \right)\right]\right) \\
    &\quad \times \Theta( L(\omega_\gamma, t, c_\theta) - R_\star )\,,
\end{aligned}
\label{eq:constr}
\end{equation}
where $ L $ is the distance the ALP traveled before its decay, see Eqs.~(3.15) in \cite{Muller:2023vjm} for the unique value $ L(\omega_\gamma, t, c_\theta) $ for any observed photon.

Using Eq.~\eqref{eq:fluence}, we can express the expected ALP-induced gamma ray flux as
\begin{equation} \label{eq:flux}
    \phi_{\gamma}=\int \diff\omega_{\gamma} \int \diff c_\theta \frac{\diff^3 F_\gamma}{\diff \omega_\gamma \diff t \diff c_\theta}\,,
\end{equation}
where the energy integration extends above the detector threshold and the angle integration covers the region satisfying the constraint in Eq.~\eqref{eq:constr}. To asses the uncertainty on this flux, we calculate it for an optimistic and a conservative choice of all relevant parameters as summarized in \cref{tab:paramValues}.
\begin{table}[t]
    \centering
    \begin{tabular}{c c c c}
        Parameter & opt. & cons. & Unit \\
        \hline
        distance $ \dSN $ & $6.70$ & $7.00$ & $ \Mpc $ \\
        mass $ M $ & 13 & 9 & $M_\odot$ \\
        radius $ R_* $ & 400 & 420 & $R_\odot$ \\
        spectral normalization $ C $ & $17.9$ & $1.36$ & $10^{48} \MeV^{-1}$ \\
        average energy $ \omega_0 $ & $87.0$ & $71.2$ & $\MeV$ \\
        spectral index $ \beta $ & $2.43$ & $2.86$ & 1 \\
        observed flux & $1.57$ & $1.47$ & $10^{-5}\cm^{-2} \s^{-1}$\\
        flux upper limit $ \Delta\phi_\gamma $ & $3.98$ & $9.10$ & $10^{-8}\cm^{-2} \s^{-1}$
    \end{tabular}
    \caption{Summary of the various parameters used to estimate the ALP-induced gamma-ray flux from SN 2023ixf. We show the range of variability that we consider to calculate the optimistic (column `opt.') and conservative (column `cons.') constraints. The references for these values can be found in the text.}
    \label{tab:paramValues}
\end{table}

\section{Gamma-ray upper limits\\ on SN 2023ixf} \label{sec:dataAnalysis}
We analyze {\it Fermi}-LAT data to assess the expected average background events $\hat{b}$ in the region around the SN explosion in a data-driven way. We select Pass 8 photon events from the 4th of August 2008 15:43:36 UTC to the 13th of June 2023 10:07:14 UTC (we denote this time interval as $T_{\rm full}$) of energies between 30 MeV and 1 TeV within a cone of radius of $15^{\circ}$ centered on the position of SN 2023ixf: $(\ell, b) = (101.9^{\circ}, 59.8^{\circ})$. The end of the chosen time period corresponds to roughly 26 days after the onset of the optical light from SN 2023ixf ({$T_{0}=$~2023-05-18~19:30:00~UTC}~\cite{2023ATel16046....1V,2023TNSAN.129....1K,2023ATel16053....1S,2023ATel16054....1S,2023ATel16057....1B,2023ATel16067....1S,2023ATel16042....1S}). To reduce the contamination of our data-set by misclassified charged cosmic-ray events, detector glitches or terrestrial gamma-ray emission, we allow for photons with zenith angle $<80^{\circ}$ to reduce the impact of photons associated with the Earth's limb and  employ the quality cuts (\texttt{DATA\_QUAL>0 \&\& LAT\_CONFIG==1 \&\& ABS(ROCK\_ANGLE)<52}). We choose the \texttt{SOURCE\_V3} event class since we require the gamma-ray signal from ALP decays to last more than $\mathcal{O}(100)$~s making it a diffuse signal rather than a transient one.

\begin{figure*}[t!]
\includegraphics[width=0.9\linewidth]{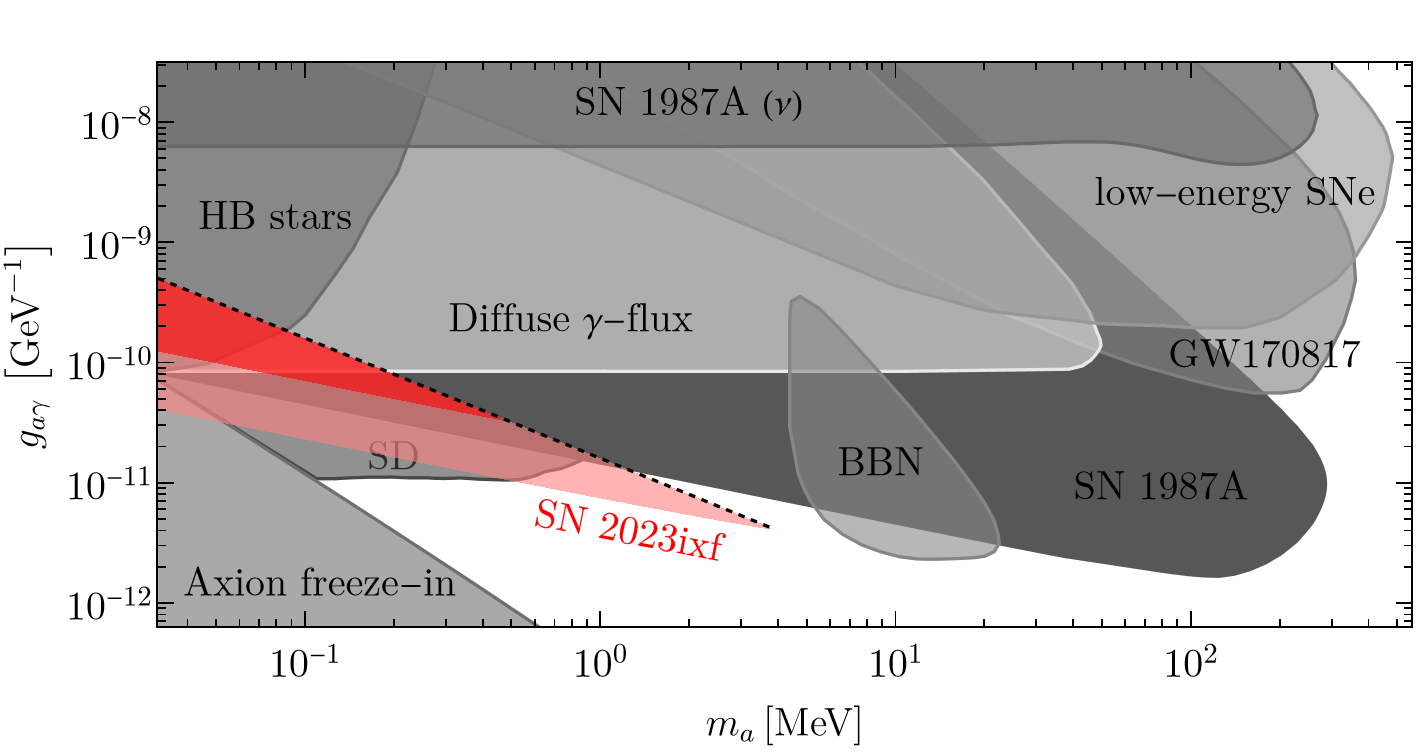}
    \caption{ALP parameter space $g_{a\gamma}-m_{a}$ with existing constraints. The constraint calculated in this work is the red exclusion region, including the optimistic (pink) and conservative (red) choice of parameters. The other shown constraints are: the cooling criterion applied to SN 1987A (SN 1987A $(\nu)$)~\cite{Lucente:2020whw}, the SN 1987A bound for decaying ALPs~\cite{Muller:2023vjm,Diamond:2023scc}, the upper limit on energy deposition by ALP-decays in the plasma of the progenitor star of SNe with a particularly low explosion energy (low-energy SNe) and the diffuse gamma-ray flux from the diffuse SN ALP background (Diffuse $\gamma$-flux)~\cite{Caputo:2022mah}, the constraint form the duration of the helium-burning phase (HB stars)~\cite{Ayala:2014pea,Carenza:2020zil,Lucente:2022wai}, the constraint from X-rays observations following GW170817 (GW170817)~\cite{Diamond:2023cto}, a bound on spectral distortions of the CMB (SD) \cite{Balazs:2022tjl}, and the irreducible cosmic ALP background from freeze-in production (Axion freeze-in)~\cite{Langhoff:2022bij}. We also show the most conservative bound from the dissociation of light elements during BBN (BBN)~\cite{Depta:2020wmr}.}
    \label{fig:bound}
\end{figure*}

It is well known that the onset of the optical light curve from a SN is not coincident with the time of core collapse. The optical signal is typically delayed. For example, for the well-known SN 1987A this delay was two to three hours~\cite{1989ARA&A..27..629A}. However, this is the only observed event and there might be a large variability in this delay time. At the moment there are no methods to estimate the delay between core-collapse and optical signal for a type II SN that are more accurate than several hours to a few days. Refs.~\cite{Meyer:2016wrm,Cowen:2010abc} use a fit of the optical signal to estimate the explosion time for a large sample of SNe, and find delay times typically around 5 days and up to 16 days. Therefore we conservatively assume that the bounce can happen at any point in the month before the optical signal. This interval will reasonably include the actual collapse event and we look for an ALP signature in this time range.
Given this uncertainty on the explosion time, a full analysis should include an averaging over all possible explosion times for any given set of ALP parameters. These determine the time evolution of the signal and it should be compared to the background during the time interval after the explosion at which the signal-to-noise ratio is maximal. However, since this procedure would be numerically costly, and as discussed below, we do not expect a large impact on the bound, we use an alternative simplified procedure here. 
We define two fiducial time intervals in which one could expect potential gamma-ray emission from SN 2023ixf: \emph{(I)} 30 days before and \emph{(II)} 7 days\footnote{As can be seen in \cref{fig:LAT_lightcurve}, there is a gap in the selected data past $ \sim 1 $ week after $ T_0 $. This region of vanishing exposure lasts for roughly 2 weeks, and hence we chose to restrict interval \emph{(II)} to the week for which data are available.} after the first optical signal (the ``ON'' time bin, for each scenario respectively). Thus, we infer the average gamma-ray background towards the position of SN 2023ixf for the fundamental time interval of $\Delta t = 30(7)$ days. We show the full light curve from the direction of SN 2023ixf for the total period in \cref{fig:LAT_lightcurve}. Time interval \emph{(I)} constrains gamma-ray signals arriving before the optical observations of the SN. This choice is made such that we can probe ALPs produced in a core collapse that happens up to a month before $ T_0 $, even if the induced signal is shorter than this period. On the other hand, interval \emph{(II)} constrains signals that last until after the onset of the optical light curve. Since the core collapse happens most likely only days, at most,  before $ T_0 $, a typical expected signal would extend from the end of \emph{(I)} into \emph{(II)}, and hence can be constrained in both time intervals.

In order to estimate the average background, we adopt the data-driven approach outlined in \cite{Muller:2023vjm}. We split the selected LAT data collected in the period $T_{\rm full}$ in temporal bins of size $\Delta t$ using the \emph{Fermi Science Tool}'s routine \texttt{gtbin} and compute the associated exposure $\varepsilon$ using \texttt{gtexposure} and the spectrum of the expected ALP-induced gamma-ray emission. Given the photon counts $N_i$  and the respective exposure $\varepsilon_i$ for all OFF bins $i$ (in the time range $T_{\rm full}$) not encompassing the SN signal, we estimate the expected background according to $\hat{b} = \alpha\bar{N}$, where $\bar{N} = \sum_i N_i$ and $\alpha = (\sum_i \varepsilon_i/\varepsilon_{\mathrm{ON}})^{-1}$ based on the assumption of Poisson-distributed photon counts. Here, $\varepsilon_{\mathrm{ON}}$ refers to the exposure of the LAT during the $\Delta t$ interval containing the SN explosion scaling $\hat{b}$ with the LAT conditions in the ON bin.
There is no significant excess above the background during either of the time intervals \emph{(I)} or \emph{(II)}, in agreement with a visual inspection of \cref{fig:LAT_lightcurve}.
Hence, we can only exclude the existence of ALPs that would lead to such an excess.

We set bounds on any type of additional gamma-ray signal $s$ in the ON region via a log-likelihood ratio test statistic based on the Poisson likelihood function
\begin{equation}
\mathcal{L}\!\left(\left.N_{\mathrm{ON}}\right|s,\hat{b}\right) = \frac{(s + \hat{b})^{N_{\mathrm{ON}}}}{(N_{\mathrm{ON}})!}e^{-(s + \hat{b})} \, ,
\end{equation}
where $N_{\mathrm{ON}}$ denotes the observed number of gamma-ray events in the respective ON scenario. Demanding that  $s\geq 0$, we set 95\% confidence level upper limits on the admissible number of additional signal photons $\hat{s}$ when the log-likelihood ratio
\begin{equation}
\lambda(s) = -2\left[\ln{\mathcal{L}\!\left(\left.N_{\mathrm{ON}}\right|s,\hat{b}\right)} - \ln{\mathcal{L}\!\left(\left.N_{\mathrm{ON}}\right|s=0,\hat{b}\right)}\right]\mathrm{,}
\end{equation}
attains a value of 2.71 as $\lambda(s)$ depends on one free parameter ($s$, or equivalently $g_{a\gamma}$ for fixed mass $m_a$) and thus follows a half-$\chi^2$-distribution with one degree of freedom \cite{2011EPJC...71.1554C}. Following this prescription, we obtain for time window \emph{(I)} $\hat{s} = 45.2$ events for an exposure of $ \varepsilon = 1.13 \times 10^{9} \cm^2 \s $ yielding an upper limit on the average flux of $ \Delta \phi_\gamma \equiv \hat s / \varepsilon = 3.98 \times 10^{-8} \cm^{-2} \s^{-1} $, while in time window \emph{(II)} we have $\hat{s} = 14.9$ events for an exposure of $ \varepsilon = 1.64 \times 10^{8} \cm^2 \s $ yielding $ \Delta \phi_\gamma = 9.10 \times 10^{-8} \cm^2 \s $. Since the resulting average flux limit is lower for interval \emph{(I)} this is shown as the optimistic case in \cref{tab:paramValues}, while interval \emph{(II)} is used for the conservative estimate.
We remark that the upper limits are set on the \emph{average flux} of the signal. Therefore, the on-set time of the signal only plays a very minor role by determining the background that the signal should be compared to, which we estimate here for the two scenarios \emph{(I)} and \emph{(II)}, and which we find to differ only by a factor $\sim 2$. Hence, the impact on the bound on $\gag$ will be marginal.

\section{Constraint from SN 2023ixf} \label{sec:constraint}
In this work, we have estimated the expected ALP induced gamma-ray signal from SN 2023ixf, \cref{eq:flux}, and compared it with observational data from {\it Fermi}-LAT. We have found no signal above background in such data, and this allows to set stringent bounds on the ALP parameters.
As discussed in \cref{sec:dataAnalysis}, in order to be conservative, we allow for a large variability in the onset of the ALP-induced signal, due to the uncertain time of core collapse in SN 2023ixf. Depending on the starting time  and duration of the ALP-induced gamma-ray signal, it might happen that it falls in a period in which the detector was not pointing to the source. For instance, in \cref{fig:LAT_lightcurve} we notice a time window in the week before $ T_0 $, characterized by a vanishing photon count for almost three days. The example signal shown as dashed red line in \cref{fig:LAT_lightcurve} falls almost completely into this region and can thus not be constraint by our analysis. Because of this limitation, we only constrain gamma-ray fluxes that do not fall below $ 90 \% $ of their maximum for three days after their onset. Longer lasting signals will, at least partially, fall into a time interval of average exposure and background flux. The other signal shown as solid red line in \cref{fig:LAT_lightcurve} is an example for this, which can clearly be excluded when compared to the measured data points and their statistical errors.
In addition, we also mention a low exposure region more than four weeks before the explosion, characterized by large error bars on the flux.

Therefore, we can set a constraint on the ALP parameter space by requiring that
\begin{equation} \label{eq:constraint}
    \left\langle \phi_\gamma(\gag, m_a) \right\rangle_{\Delta t} < \Delta \phi_\gamma \equiv \frac{\hat s}{\varepsilon} \, ,
\end{equation}
i.e.~that the average ALP-induced photon flux in the energy range $ 30 \MeV $ to $ 1 \TeV $ during the time interval $ \Delta t $ is smaller than the upper limit on the flux $ \Delta \phi_\gamma $. The latter is defined as the additional number of signal photons $ \hat s $ allowed at the 95\% confidence level, as derived in the previous section, divided by the exposure $ \varepsilon $ during the time interval $ \Delta t $.
As discussed above and summarized in \cref{tab:paramValues}, we derive our bounds in an optimistic and a conservative case, with different parameter values determining the ALP-induced gamma-ray flux, as well as different flux upper limits corresponding to time intervals \emph{(I)} and \emph{(II)}, respectively.
Note that the intervals \emph{(I)} and \emph{(II)} are only examples that determine a background and an observation duration, and it is not necessary for the signal to actually have occurred during these exact periods. As can be seen from \cref{eq:constraint}, we compare an \emph{average flux} with the data. When exactly this average flux would have been observed only determines which data to compare to and could determine a more appropriate averaging window $\Delta t$ as the signal might be somewhat shorter than our fiducial time intervals. None of these effects are expected to be very important numerically.
The result is shown in \cref{fig:bound}, where the excluded parameter region is shown as an opaque red region for the conservative case and as a transparent pink region for the optimistic case. Note that in the excluded parameter region fireball formation as discussed in \cite{Diamond:2023scc} does not occur.
The conservative case falls in a region excluded by other stellar arguments, confirming the existing constraints.
In the optimistic case, the resulting bound is stronger than that from the same non-observation of an ALP-induced gamma-ray signal following SN 1987A. This is a remarkable result because SN 1987A exploded more than 100 times closer to Earth compared to SN 2023ixf, and hence the expected flux was a factor of $ \sim 10^4 $ higher.
We note that the largest difference between optimistic and conservative case is the ALP spectrum, while the upper limit on the gamma-ray flux is only a factor of $ ~2 $ different. Thus, also with the conservative flux limit, but optimistic assumptions about the ALP spectrum, the resulting bound would be stronger than the previous one from SN 1987A.\footnote{It is worth to mention that since the observation of the supernova, and also since this work was first made public there has been an intense activity in determining the mass of the SN 2023ixf progenitor~\cite{VanDyk:2023tns,Niu:2023tmz,Jencson:2023bxz,Neustadt:2023fao,2023TNSAN.139....1S,Pledger:2023ick,Kilpatrick:2023pse}. Currently, the largest viable mass range is $9-22$~$M_{\odot}$, depending on the analysis. This shows that our choice of the parameter range is quite representative of the current state of knowledge and in agreement with the majority of these analyses. Most of the uncertainty on the proposed constraint is associated with this parameter.}

We do not constrain ALP parameters above the black dashed line in \cref{fig:bound}, as argued in \cref{sec:constraint}, since the signal would be varying on time scales shorter than three days. However, in principle, a constraint can be derived for these ALP models as well, if additional assumptions about the time of core collapse are made. Then, the onset of the gamma-ray signal would be known, and in case there is an overlap between a non-zero exposure of \lat~and the signal, the underlying ALP parameters might be excluded. Similarly, even the bound below the black dashed line could be modified if the core collapse time was known and the time-dependence of the signal could be taken into account. In this case, \cref{eq:constraint} could be applied only in the time period during which the signal has its peak. This would slightly strengthen the bound if the signal peak is either shorter or much longer than the 30 or 7 days considered here -- provided that the flux upper limit is comparable to those assumed in our cases \emph{(I)} or \emph{(II)}. However, we expect a relatively small improvement of our optimistic and conservative bounds once the core-collapse time of SN 2023ixf is known because the variations of the measured photon count and its error are not very large (outside of the zero-exposure regions, see \cref{fig:LAT_lightcurve}), the signal peaks are not much shorter than our conservative $ \Delta t = 7 $~days for the relevant ALP parameters, and because the bound on the ALP-photon coupling scales as $ \gag \sim \phi_\gamma^{1/4} $. Nevertheless, it would be an interesting extension of this work to take the time-dependence of the signal into account, especially because it might extend the bound to larger masses and couplings.

\section{Conclusions}
\label{sec:conclusions}
This analysis shows the importance of an accurate characterization of astrophysical signatures of ALPs and, more generally, exotic particles. It is well-known that a future Galactic SN will be a revolutionary opportunity to deepen our knowledge of fundamental physics. However, as we demonstrated here, and in line with similar conclusions for light ALPs~\cite{Crnogorcevic:2021wyj}, also the more frequent extragalactic SNe can probe unexplored parameter regions of heavy ALPs.
The limited information on the SN core collapse time reflects in large uncertainties in searches for light ALPs because their signatures are expected in a coincidence of just a few seconds after the core-collapse. This feature makes these studies very sensitive to the timing information. By contrast, massive ALPs give rise to a signal persistent in time, more similar to a steady source. As argued in this work, this allows us to be much less sensitive to the uncertainty on the time of the  collapse of the core. Hence, core-collapse SNe in near-by galaxies are promising laboratories for new physics. We expect this study to be the first of many searching for the characteristic signal induced by the decay of exotic particles from this and other extragalactic SNe. They present an excellent opportunity to study fundamental physics using naturally occurring astrophysical phenomena that should be exploited.

\begin{acknowledgements}
We thank Francesca Calore, Maurizio Giannotti, Giuseppe Lucente, Alessandro Mirizzi, and Edoardo Vitagliano for helpful discussions.
For this work, we have made use of the \textit{Axion-Limits} repository \cite{AxionLimits}.
The work of E.~R.~and P.~C.~is supported by the European Research Council under Grant No.~742104 and by the Swedish Research Council (VR) under grants  2018-03641 and 2019-02337. The work of C.~E.~is supported by the ``Agence Nationale de la Recherche'' through grant ANR-19-CE31-0005-01 (PI: F.~Calore), and has been supported by the EOSC Future project which is co-funded by the European Union Horizon Programme call INFRAEOSC-03-2020, Grant Agreement 101017536. A.~G.~acknowledges support from the Swedish Research Council, Dnr 2020-03444.
This article/publication is based upon work from COST Action COSMIC WISPers CA21106, supported by COST (European Cooperation in Science and Technology).
\end{acknowledgements}

\bibliographystyle{bibi}
\bibliography{biblio.bib}

\end{document}